\documentstyle[12pt,aaspp4,here,flushrt]{article}
\begin{document}
\title{Uncertainties in the Determination of Primordial D/H} 
\author{Antoinette Songaila\altaffilmark{1}}

\affil{Institute for Astronomy, 2680 Woodlawn Drive, Honolulu, HI 96822, USA}

\altaffiltext{1}{The author was a visiting astronomer at the W. M. Keck
  Observatory, jointly operated by the California Institute of Technology and
  the University of California.}


\begin{abstract}
The current status of high redshift D/H measurements is discussed.  
I first examine whether the observations of HS1937$-$1009 require a 
low value of D/H.  It is shown that the LRIS measurements of the 
continuum break and the high resolution Lyman series data can easily 
be modelled with D/H in the range  $5 \times 10^{-5}$ to $10^{-4}$. 
I then discuss measurements on weaker Lyman limit systems.
A statistical treatment of 10 partial Lyman limit systems favors 
$5 \times 10^{-5} \le {\rm D/H} \le 2 \times 10^{-4}$.
\end{abstract}
\section{Introduction}
The advent of the 10 meter telescopes has made possible the important
observational search for the value of primordial D/H from high redshift
quasar absorption lines (\cite{SCHR}) which in principle is a sensitive probe of
$\Omega_{\rm baryon}$\ and is ultimately a diagnostic of the health of the
standard big bang nucleosynthesis (SBBN) picture.  Unfortunately, serious
underestimates of the systematic difficulties in such measurements have led to
overenthusiastic claims of the precision of both high (\cite{HR}) and low
(\cite{TFB}) (TFB) values, with interesting repercussions for the theoretical
picture, particularly as the original low value of (\cite{TFB}) would have been
seriously incompatible with measurements of $\Omega_{\rm baryon}$\ from the
other light elements, notably $^4{\rm He}$\ (\cite{olive}).  However, the
prospect of an observational incompatibility with SBBN has had the stimulataing
effect of provoking a scrutiny of the subject, that has for instance cast serious
doubt on the D + $^3$He argument and has even provoked the abandoning of the
primary dependence of SBBN in determining $\Omega_{\rm baryon}$\ in favor of
non-nucleosynthetic constraints on the cosmological parameters (\cite{steigman}).

In the long run, what are the prospects for obtaining a believable value for
primordial D/H from quasar absorption line measurements and, perhaps more
importantly, demonstrating or convincingly ruling out any variability?  It is my
opinion that {\it only\/} the accumulation of measurements from many systems
will provide either of these answers.  Considerable attention has been paid to
the possibility of the chance contamination of the high values (\cite{TBK}) by
intervening Lyman forest material, a systematic problem in the use of partial
Lyman limit systems, for which reason their use must {\it always\/} be
statistical only (\cite{SCHR}).  
In section 3 I will discuss my own current sample of observations of 10 PLLS
quasars and argue that it favors the range $5 \times 10^{-5} \le {\rm D/H} \le 2
\times 10^{-4}$.  However, I will also attempt to show (section 2) that the
alternative strategy of using higher column density systems is also ultimately
susceptible to yielding (lower) limits {\it only\/}, this despite the best
efforts to model the systematic effects (\cite{SWC}) (SWC);\cite{BT} (BT)).  

Fortunately, some very nice measurements of other systems are beginning to
appear (\cite{Webb};\cite{Vid}) as John Webb and Alfred Vidal-Madjar discuss
here.  However, I would caution that the high D/H value at low redshift is
also subject to the large systematic uncertainty from chance contamination by
the internal hydrogen structure of the system and that it is perhaps premature
to assume from this one measurement that there is variation in D/H among
quasar lines of sight.

\section{D/H toward HS1937$-$1009}
There has been much controversy surrounding TFB's low value (\cite{TFB}) of
${\rm D/H} = 2.3 \pm 0.3 \pm 0.3 \times 10^{-5}$\ from the $z = 3.572$\ Lyman
limit system toward HS1937$-$1009, questioning the claimed small systematic
errors in view of both the considerable uncertainty in the total H~I column
density in this system and in the lack of uniqueness of the assumed cloud
distribution.  Wampler
(\cite{Wamp}) showed that simple cloud models could give D/H as high as
$6.3\times 10^{-5}$\ and still be compatible with TFB's fit.  SWC measured the
total H~I column density from LRIS observations of the Lyman continuum break
and even with their most conservative modeling of the continuum they revised
the TFB H~I value downward to $5.9\times 10^{17}~{\rm cm}^{-2}$ (\cite{SWC}),
outside the TFB published errors (\cite{TFB}), and with more realistic modeling
of the continuum preferred a value of about $5\times 10^{17}~{\rm cm}^{-2}$.
BT subsequently also revised the H~I column density downward based on similar
low resolution spectroscopy but using a HIRES spectrum to remove the effects
of the forest above the break in an LRIS spectrum; in this way they derived a
higher value of $7.24\pm 0.35 \times 10^{17}~{\rm cm}^{-2}$ (\cite{BT}).  This
procedure is problematical because, while the forest can be deblended above
the break from HIRES data (subject to uncertainties in continuum fitting) it
can only be modeled below the break.  Unless the models are a perfect
representation of the forest there is a substantial possibility of introducing
matching error.  The more direct process used by SWC avoids this by simply
assuming continuity of the forest properties across the break.

Figure~1 shows a 3.4 hr LRIS spectrum of the quasar HS1937$-$1009.  The
spectrum has been smoothed to 50~\AA\ resolution and the effects of the $z =
3.572$\ Lyman limit system itself have been divided out for the two values of
$N({\rm H~I}) = 8.9\times 10^{17}~{\rm cm}^{-2}$\ 
given by TFB (\cite{TFB}) and of $N({\rm H~I}) = 7.24\times 10^{17}~{\rm cm}^{-2}$\ 
given by BT (\cite{BT}) .  The solid line shows
the power law fit to the quasar spectrum, in the absence of Ly$\alpha$\ forest
absorption, given by (\cite{Zheng}).  The effect of the forest lines
produces the deep decrement between the observed spectrum and the continuum
level which is, however, quite smoothly varying at this resolution.  Note that
the peak at 4900~\AA\ is the O~VI emission line.  The dashed line marks the
continuum edge of the Lyman limit system.  The TFB value is completely
unphysical, in that it requires negative forest absorption.  The BT value 
is also
problematic in that it requires an abrupt drop in the forest opacity just at
the position of the continuum edge, suggesting that there is indeed a mismatch
between the modeling below the break and the direct deblending of the forest
above it.  A value of $N({\rm H~I}) \sim 6\times 10^{17}~{\rm cm}^{-2}$\
produces a smooth forest opacity above and below the break though, as argued
in SWC, even this may be an overestimate of $N({\rm H~I})$\ because the forest
opacity is expected to increase at these wavelengths (\cite{SWC}).


However, even if one adopted the upper-bound BT value of 
$N({\rm H~I}) = 7.59 \times 10^{17}~{\rm cm}^{-2}$\ the
classical problems of fitting saturated lines leave a
very large uncertainty in deriving the D/H value, as was
emphasized by Wampler (\cite{Wamp}).  The original TFB cloud
model gives ${\rm D/H} = 3.0
\times 10^{-5}$\ for $N({\rm H~I}) = 7.24 \times 10^{17}~{\rm cm}^{-2}$\
and ${\rm D/H} = 4.0
\times 10^{-5}$\ for $N({\rm H~I}) = 6 \times 10^{17}~{\rm cm}^{-2}$.
The latter value matches SWC's estimate (\cite{SWC}) of a reasonable minimum D/H
ratio in this system.  However, in Figure~2 we show a direct comparison of a
model with $N({\rm H~I}) = 7.5 \times 10^{17}~{\rm cm}^{-2}$\ and ${\rm D/H} =
5\times 10^{-5}$ with the original TFB model having $N({\rm H~I}) = 8.9 \times
10^{17}~{\rm cm}^{-2}$.  The new model actually gives a formally better fit to
a new set of HIRES observations of the quasar, but within the systematic
errors both are probably adequate descriptions.  With a lower, and as we have
argued above, more realistic value of $N({\rm H~I})$ a D/H near $10^{-4}$ can
also be accommodated, as shown in the second set of panels.  The fundamental
result here is simply the difficulty in fitting the saturated lines in strong
Lyman limit systems and the corresponding uncertainties in determining D/H in
such systems.  However, it should be noted that a $2~\sigma$\ upper bound of
${\rm D/H} \le 3\times 10^{-5}$, required for consistency with $\eta_{10} \ge
6$\ derived from non-BBN constraints (\cite{steigman}), would be difficult to
accommodate with any proposed current value of $N({\rm H~I})$.
\section{Partial Lyman Limit Systems}
The use of weaker Lyman limit systems avoids many of these problems
because the high order Lyman series lines desaturate, which permits
a direct and unambiguous measurement of the hydrogen kinematic structure.
However, because of the lower $N({\rm H~I})$ the D is weaker and 
more susceptible to chance contamination by floating H~I lines.
Such contamination may be either from random forest lines or from the
internal kinematic structure of the Lyman limit system itself.  The
latter contamination is harder to deal with and persists even in low redshift
objects.

In order to approach this problem in a robust way I have used my current 
sample of HIRES observations of ten quasars with Lyman limit systems
having $5\times 10^{16} \le N({\rm H~I}) \le 5\times 10^{17}\ 
{\rm cm}^{-2}$.  In each case I established the kinematic structure
of the H~I and then determined how much D~I could be present at both
the true position and also at a position at +82 km~s$^{-1}$ which should,
on grounds of symmetry, 
be on average indistinguishable from the D position if there were no
significant deuterium present.

The upper limits on the D/H  and false D/H  ratios were
calculated under both turbulent broadening and thermal broadening 
assumptions and are shown in Figure~3 for the more constraining
thermal broadening case.  Four of the ten systems have relatively
weak absorption at the false red position, but in only one case
is ${\rm D/H} < 1.4 \times 10^{-4}$ obtained at the true position.
(This rises to $2 \times 10^{-4}$ for turbulent broadening.)\ \
A rank sum test shows that the two distributions are
inconsistent at the 95\% confidence level, which is a rather
marginal result, but does favor having ${\rm D/H} > 5 \times 10^{-5}$ to
bring the two sides into consistency.  Together with the measured upper
bounds this would give $5 \times 10^{-5} < {\rm D/H} < 1.4 \times 10^{-4}$
for thermal broadening.

\acknowledgements{This work was supported by NSF grant AST 96-17216.}


\newpage

\begin{figure}
\plotone{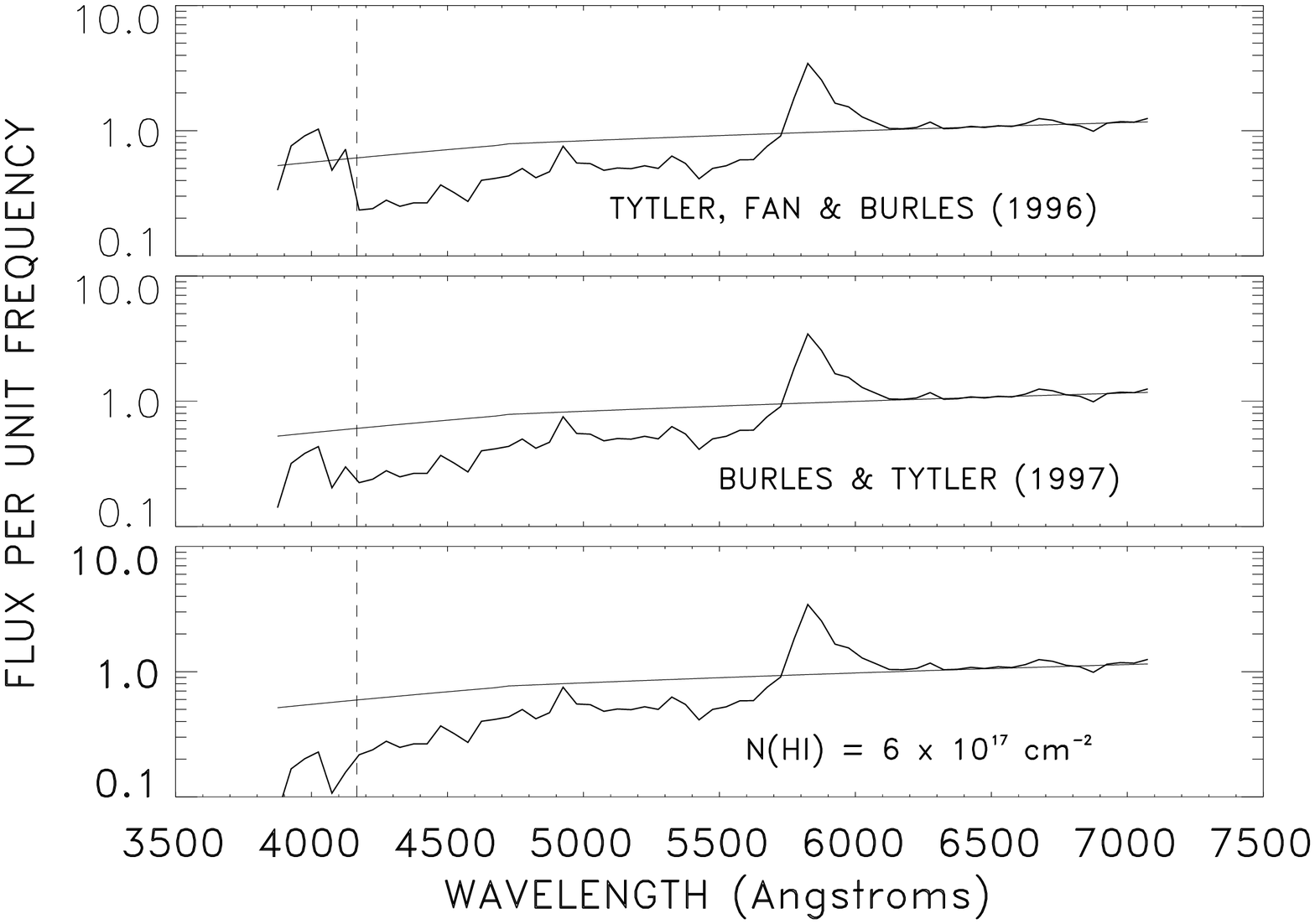}
\caption{The effect of different assumed values of neutral hydrogen column
density on the Lyman continuum break region of the spectrum of HS1937$-$1009
with the effect of the Lyman limit system divided out.  See text for details.
}
\end{figure}

\begin{figure}
\plotfiddle{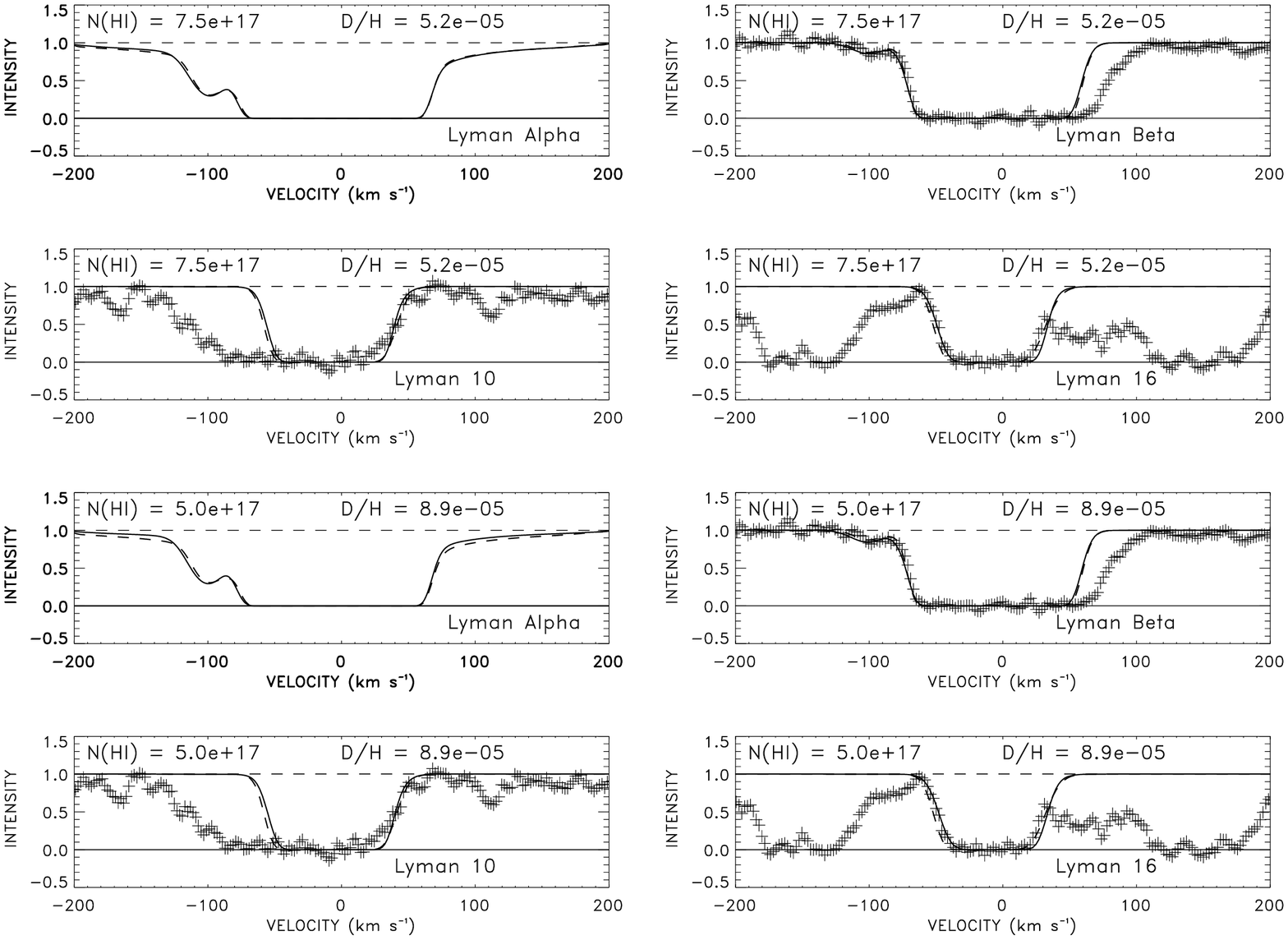}{7in}{0.}{60}{110}{-216}{0}
\caption{
Comparison between the cloud model of TFB (dashed line) and (upper
panels) one with ${\rm D/H} = 5.2\times 10^{-5}$\ for the $1~\sigma$\ upper
bound on the H~I column density of $7.59 \times 10^{17}~{\rm cm}^{-2}$\
claimed in Burles \& Tytler ; and (lower panels) for a model
with $N({\rm H~I}) = 5.0 \times 10^{17}~{\rm cm}^{-2}$\ (SWC's
favored value) and ${\rm D/H} = 8.9 \times 10^{-5}$.  
}
\end{figure}

\begin{figure}
\plotone{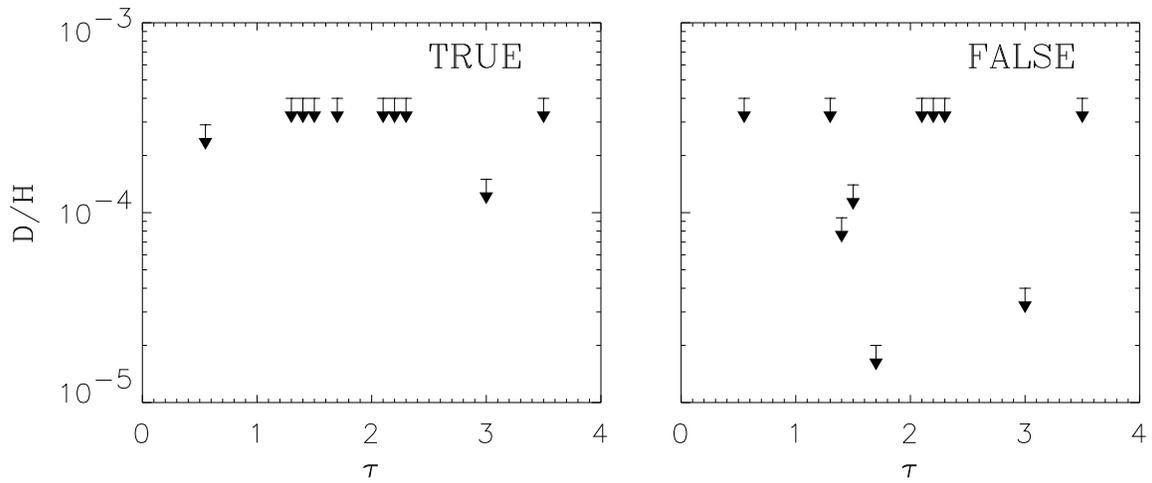}
\caption{Upper bounds on D/H determined for 10 partial Lyman
limit systems are shown as a function of the optical depth of the
Lyman limit system ({\it left panel}).  The same measurement carried
out at the symmetrically placed redward position gives much tighter
constraints ({\it right panel}), suggesting that there is indeed
enhanced absorption caused by deuterium in the blue wing.}
\end{figure}


\begin{thebibliography}{}

\bibitem[Burles \& Tytler 1997]{BT} Burles, S. \& Tytler, D. 1997, AJ, in
press [astro-ph/9707176]
\bibitem[Olive et al.\ 1997]{olive} Olive, K. A., Skillman, E., \& Steigman, G. 1997, ApJ, 483, 788
\bibitem[Rugers \& Hogan 1996]{HR} Rugers, M. \& Hogan, C. J. 1996, AJ, 111, 2135
\bibitem[Songaila et al.\ 1994]{SCHR} Songaila, A., Cowie, L. L., Hogan, C. \&
Rugers, M. 1994, Nature, 368, 599
\bibitem[Songaila, Wampler \& Cowie 1997]{SWC} Songaila, A., Wampler, E. J.,
\& Cowie, L. L. 1997, Nature, 385, 137
\bibitem[Steigman et al.\ 1997]{steigman} Steigman, G., Hata, N., \& Felten,
J. E. 1997, preprint [astro-ph/9708016] 
\bibitem[Tytler, Fan \& Burles 1996]{TFB} Tytler, D., Fan, X.-M., \& Burles,
S. 1996, Nature, 381, 207
\bibitem[Tytler, Burles \& Kirkman 1997]{TBK} Tytler, D., Burles, S., \&
Kirkman, D. 1997, submitted to ApJ [astro-ph/9612121]
\bibitem[Vidal-Madjar et al.\ 1997]{Vid} Vidal-Madjar, A. et al. 1997, these
 proceedings
\bibitem[Wampler 1996]{Wamp} Wampler, E. J. 1996, Nature, 383, 308
\bibitem[Webb et al.\ 1997]{Webb} Webb, J. et al. 1997, these proceedings 
\bibitem[Zheng et al.\ 1997]{Zheng} Zheng, W. et al.\ 1997, ApJ, 475, 469

\end{thebibliography}
\end{document}